\documentstyle[12pt,aaspp4,epsfig]{article}

\slugcomment{To appear in {\it The Astrophysical Journal}}
\received{}
\accepted{}
\journalid{}{}
\articleid{}{}

\lefthead{Halpern et al. }
\righthead{}

\begin{document}

\def\source{3EG~J2016+3657}
\def\radio{B2013+370}
\def\ro{{\it ROSAT\/}}
\def\asca{{\it ASCA\/}}

\title{3EG J2016+3657: Confirming an EGRET Blazar Behind the Galactic Plane}

\author {J. P. Halpern\altaffilmark{1,2}, M. Eracleous\altaffilmark{2,3},
R. Mukherjee\altaffilmark{4}, E. V. Gotthelf\altaffilmark{1}}

\altaffiltext{1}
{Columbia Astrophysics Laboratory, Columbia University, New York, NY 10027}
\altaffiltext{2}
{Visiting Astronomer, Kitt Peak National Observatory,
National Optical Astronomy Observatories, which is
operated by the Association of Universities for
Research in Astronomy, Inc. (AURA) under cooperative
agreement with the National Science Foundation.}
\altaffiltext{3}
{Dept. of Astronomy \& Astrophysics, The Pennsylvania State University,
University Park, PA 16802}
\altaffiltext{4}
{Dept. of Physics \& Astronomy, Barnard College \& Columbia University, New York, NY 10027}

\bigskip

\begin{abstract}

We recently identified the blazar-like radio source G74.87+1.22
(B2013+370) as the counterpart of the high-energy $\gamma$-ray source
\source\ in the Galactic plane.  However, since most blazar identifications
of EGRET sources are only probabilistic in quality even at high Galactic
latitude, and since there also exists a population of unidentified
Galactic EGRET sources, we sought to obtain additional evidence to
support our assertion that \source\ is a blazar.  These new observations
consist of a complete set of classifications for the 14 brightest
\ro\ X-ray sources in the error circle, of which B2013+370 remains
the most likely source of the $\gamma$-rays.  We also obtained further
optical photometry of B2013+370 itself which shows that it is variable,
providing additional evidence of its blazar nature.  
Interestingly, this field contains, in addition to the blazar,
the plerionic supernova remnant CTB 87, which is too distant to be the
EGRET source, and three newly discovered cataclysmic variables,
all five of these X-ray sources falling within $16^{\prime}$ of each other.
This illustrates the daunting problem of obtaining
complete identifications of EGRET sources in the Galactic plane.

\end{abstract}


\section{Introduction}

In a recent multiwavelength study of the region in Cygnus containing
the unidentified COS--B $\gamma$-ray source 2CG~075+00, Mukherjee et al.
(2000) noted that two discrete EGRET sources, 3EG J2021+3716 and
\source, are consistent with the COS--B location (Pollack et al. 1985),
and that the weaker of these two, \source, is coincident with a 
blazar-like radio source \radio\ (G74.87+1.22) which they proposed 
as its most likely identification.  The temporal variability and 
broad-band spectral properties
of \source\ are consistent with those of other EGRET blazars.
Leaving aside the novelty of a blazar located behind the Galactic plane,
the identification of any particular EGRET source with a blazar is
generally a probabilistic claim, since only a small fraction of known blazars
were seen to be active in $\gamma$-rays during the EGRET survey.
Furthermore, there is clearly a Galactic population of $\gamma$-ray
sources at low latitude that are mostly unidentified.
The {\it a posteriori} probability that the blazar \radio\ is the 
correct identification of the EGRET source 3EG J2016+3657 was estimated 
following the method of Mattox et al. (1997) as $\approx 98.8\%$,
even though the {\it a priori} probability that EGRET will detect a 
random radio source having the properties of \radio\ is only 5.8\%.
[While the distinction between these two types of probability should
be clear, they are sometimes confused, causing serious
mistakes to be made in diverse fields of inquiry,
as was illustrated by Good (1995).]

In this particular case, the reliability
of the identification of \radio\ with \source\ is as good as that
of the well-identified EGRET blazars listed by Mattox et al. (1997), but it
is slightly diminished by the location of \source\ in the Galactic plane,
where an increased density of $\gamma$-ray sources resides
which have proven even more difficult
to identify than the blazars.  One way to further test the association of 
\radio\ with \source\ is to conduct a deep search for plausible 
alternative $\gamma$-ray source counterparts within the error circle.
In the case of \source, we report such an investigation here which consists 
of complete optical spectroscopic identifications of all soft and hard
X-ray sources in the vicinity of \source\ to faint limits which,
by process of elimination, leaves \radio\ as the most likely
counterpart of \source. 

\section{X-ray Observations}

The error circle of \source\ was covered by several imaging X-ray
observations, including the {\it Einstein} IPC (Wilson 1980), the
\ro\ PSPC, the \ro\ HRI, and the \asca\ GIS.  The
results of these observations were described
by Mukherjee et al. (2000). In Figures 1 and 2,
we show the \ro\ PSPC and
HRI images with X-ray sources numbered as in that paper.  Table
1 gives positions of the brightest sources in the HRI image,
or in the PSPC in the case of sources not covered by the HRI.  
The uncertainties in position are statistical only, and do not
include any systematic offsets (see below).  The
limiting HRI count rate of $\approx 1.4 \times 10^{-3}$~s$^{-1}$
corresponds to an unabsorbed 0.1--2.4~keV
flux of $6.8 \times 10^{-14}$ ergs cm$^{-2}$ s$^{-1}$
in the case of a thermal plasma of temperature $T = 3 \times 10^6$~K
and $N_{\rm H} = 3 \times 10^{20}$~cm$^{-2}$ as might be appropriate
for stellar coronal sources, or $9.5 \times 10^{-13}$ ergs cm$^{-2}$ s$^{-1}$
in the case of a $\Gamma = 2$ power law and the total Galactic
$N_{\rm H}$ of $ 1.3 \times 10^{22}$~cm$^{-2}$, which might apply to distant
Galactic or extragalactic sources.

\section{Optical Observations}

Of the 14 X-ray sources in Table~1, eight were identified in
Mukherjee et al. (2000). We were able to complete the identifications 
either from positional coincidences with bright stars (Figure 3) or
by obtaining spectra of the nearest optical object to the X-ray position
using CCD spectrographs on either the MDM 2.4m telescope or the KPNO 2.1m 
telescope in 2000 June.  The only source with no optical counterpart
is the well-known supernova remnant CTB 87.
Figure~4 shows CCD images from the 2.4m telescope that can serve
as finding charts for the fainter counterparts, and Figure~5 shows
our collection of optical spectra.
Table 1 lists R magnitudes either measured from our CCD images for
the faint objects, or from the USNO--A2.0 catalog 
(Monet et al. 1996) for the brighter
stars.  Optical positions were also taken from the USNO--A2.0 or
from the SIMBAD data base.  The agreement between optical and
X-ray positions is excellent.  Figure~6 illustrates the offsets
between X-ray and optical positions.  There is evidently 
a systematic error in the HRI aspect solution of
$\approx 4^{\prime\prime}$, which accounts for the
tight group of sources displaced from the origin
with small statistical errors.  Such a systematic error
is typical for \ro\ and not of concern here.

\section{Comments on Individual Sources}

Entries in Table~1 and in Figures 1 and 2 follow the numbering
scheme in Mukherjee et al. (2000).  In this section, we give details
on each of these sources using the same identifying numbers.

{\it 1. CTB 87\/}: This well-studied supernova remnant,
also known as G74.9+1.2,
is an extended radio source with a flat spectrum, filled center,
and high polarization.  It is comparable to the Crab in its radio
properties (Duin et al. 1975; Weiler \& Shaver 1978; Wilson 1980),
but it is characterized as a ``non Crab-like plerion'' by
Woltjer et al. (1997), who enumerate a class which is
weak in X-rays relative to radio
and lack an observed pulsar, at least so far.
Its H~I absorption spectrum indicates a distance of 12~kpc.
An association of a SNR with an EGRET source can be hypothesized either
due to an embedded $\gamma$-ray pulsar, of which several are known, or
to the decay of $\pi^0$'s created in the SNR shock
(Montmerle 1979; Aharonian et al. 1994), a theory to
account for unidentified EGRET sources which has yet to be verified
observationally.

Wilson (1980) argued that the X-ray luminosity of CTB 87,
which is 100 times less than that of the Crab, implies that the spin-down
power of the embedded pulsar must be correspondingly less,
$I\Omega\dot \Omega \sim 4 \times 10^{36}\ (d/12\ {\rm kpc})^2$~ergs~s$^{-1}$.
It would require a factor of 2 more power 
than this to be emitted in the EGRET energy
band alone to account for the EGRET measured flux of
$\simeq 5 \times 10^{-10}$ ergs~cm$^{-2}$~s$^{-1}$ from 3EG~J2016+3657.
This argument is insensitive to distance as long as the
X-ray synchrotron nebula is considered a calorimeter of the present
pulsar power.  Alternatively, Gaensler et al. (2000) suggested that the
spin-down power of the pulsar in G74.9+1.2 is much larger, 
$\sim 1.8 \times 10^{38}$~ergs~s$^{-1}$ based on the similarity of its
radio luminosity to that of the Crab.  These differences of opinion
attest to our incomplete understanding of the physics of pulsar
synchrotron nebulae.  For the purposes of this investigation, it matters
little because the Crab pulsar itself channels only about 0.2\% of
its spin-down power into $\gamma$-rays.   Even the larger
power estimate of Gaensler et al. (2000) is less than half
the Crab spin-down power, and the distance is 6 times greater.
The flux from such a pulsar would then be at most 1/70 times
that of the Crab.  EGRET would not detect
such a source, especially in the confusing Cygnus region.

If, instead, CTB 87 hosts a Geminga-like pulsar whose
energy is no longer trapped by the nebula, and is maximally efficient
in the production of $\gamma$-rays, then we would expect a spin-down
power of only $\sim 3 \times 10^{34}$~ergs~s$^{-1}$.  Such a pulsar
is inadequate to explain the flux of 3EG~J2016+3657 unless it were
at $d < 500$~pc, which is certainly ruled out by the H~I and X-ray
measured column density to the SNR.   Thus, the remnant CTB 87 is
unlikely to be responsible for the EGRET source 3EG J2016+3657.
It is doubtful that any pulsar at a distance of 12~kpc was
detected by EGRET.

\medskip

{\it 2. RX J2015.6+2711\/}:  This source has the optical spectrum
of a cataclysmic variable (CV), probably of the magnetic type since
its He~II~$\lambda$4686 emission line is as strong as
H$\beta\ \lambda$4861 (see Figure 5).
The optical spectrum is clearly reddened, so it may
be more distant than 1~kpc.  There is no reason to suppose that
this, or any other cataclysmic variable, is the source of
\source, as none of the dozens of nearby CVs have been detected
by EGRET (Barrett et al. 1995; Schlegel et al. 1995).

\medskip

{\it 3. B2013+370\/}: 
The 2~Jy radio source \radio\ (G74.87+1.22)
is a well-studied compact, flat-spectrum radio source that was 
first noticed during the study of the SNR CTB 87 (Duin et al. 1975). 
Its multiwavelength properties were compiled by Mukherjee et al. (2000),
who noted, in agreement with previous authors, that it has the standard
radio properties of a blazar, but now with the addition of optical and 
X-ray evidence.  Although the Galactic extinction in this direction
is considerable, $E(B-V) =1.82$ mag (Schlegel et al. 1998), we were able
to get excellent $R$ band images of it using the MDM 2.4m telescope
on 2000 April 24, and again on 2000 July 17.  The seeing on both occasions
was $\approx 0^{\prime\prime}\!.75$.  An unresolved object appears in these
images less than $1^{\prime\prime}$ from the VLBI radio position
(Duin et al, 1975), and it is clearly variable as shown in Figure~4.
Using Landolt (1992) standard stars,
we measure calibrated magnitudes of $R = 21.40 \pm 0.04$ on April 24,
and $R = 21.82 \pm 0.05$ on July 17.  [A preliminary magnitude of
$R = 21.6 \pm 0.2$ was reported by Mukherjee et al. (2000)
for the April image,
which at that time was still uncalibrated].  After correcting for Galactic
absorption, these magnitudes become $R = 16.53$ and $R = 16.95$, respectively.
These results are in accord with the multiwavelength spectral 
and variability properties of typical blazars.

We note that this object was also probably detected in the $I$ band
by Geldzahler et al. (1984), who found $I = 19.5 \pm 0.5$, while
also noting that it appeared extended.  Our $R$-band images show that 
the object closest to the radio position is unresolved and variable,
while a faint star $1.^{\prime\prime}7$ to the northwest of it was likely
responsible for the extended appearance in the Geldzahler et al.
image.

Mukherjee et al. (2000) argued that \radio\ has all the characteristics of 
a compact, extragalactic, non-thermal radio source that is typical of the 
many extragalactic sources seen by EGRET.  The reader is referred to that
paper for the details.  The detection of optical variability reported
here bolsters those arguments.  Although there is only weak evidence
for gamma-ray variability from \source, the data allow variability
by at least a factor of two, which is greater than the optical
amplitude seen so far.

\medskip

{\it 4. RX J2015.4+2711\/}:  This source appears to be a garden
variety CV with strong emission lines of H and He~I,
and is probably the closest of the three CVs discovered in this
study because it is the least reddened.

\medskip

{\it 5. RX J2015.6+2704\/}:  This is another reddened CV,
with emission lines of H and weak He~I.  It is interesting
to note that the expected number of CVs
in one \ro\ HRI field in the Galactic plane is $\approx 1$
assuming a local space density of $6 \times 10^{-6}$~pc$^{-3}$
(Patterson 1984), and that in the Galactic plane \ro\ can detect them to a
distance of $\sim 2$~kpc.  Therefore, the probability of
detecting at least three CVs in one field is not so small, $\approx 0.20$.
Based on a survey of {\it Einstein} X-ray sources in the Galactic
plane, Hertz et al. (1990) claimed that the space density of CVs is
several times higher, $\simeq (2-3) \times 10^{-5}$~pc$^{-3}$,
which would also be consistent with our result.

\medskip

{\it 6. HD 228766\/}: Originally classified as WN7,
this is actually a binary O star system (Hiltner 1951;
Walborn 1973) with a period of 10.74~d, consisting of an O7.5
primary and an O5.5f secondary (Massey \& Conti 1977).  A minimum
mass of $16\,M_{\odot}$ for each star was derived by Massey \& Conti,
who also estimated a mass loss rate of $10^{-5}\,M_{\odot}$~yr$^{-1}$
from
the Of star.  Such high mass loss rates are of interest in relation
to the theory of White \& Chen (1992) which predicts $\pi^0$-decay
$\gamma$-rays of as much as $10^{35}$~ergs~s$^{-1}$
from shocks in the densest parts of the stellar wind.
However, at the distance of 3.5~kpc
(Chlebowski, Harnden, \& Sciortino 1989), such a
flux from HD 228766 is not likely to be detectable by EGRET.
It is variable between two X-ray observations separated by
5 months.

\medskip

{\it 7. HD 193077\/}: Also cataloged as WR 138
(van der Hucht et al. 1981), this star is of subtype WN5+OB. 
A probable binary period of $\simeq 4.2$~yr was measured by Annuk (1990).
It is thought to
be a member of the Cyg OB1 association at $d = 1800$~pc (Hamann, Koesterke,
\& Wessolowski 1993).  These authors also derived a mass loss rate of
$2.5 \times 10^{-5}\,M_{\odot}$~yr$^{-1}$ and a terminal wind velocity
of 1500~km~s$^{-1}$.
The association of WR 138 with \source\ was suggested by Kaul \& Mitra (1997),
and by Romero, Benaglia, \& Torres (1999).  As the brightest 
\ro\ X-ray source in the field of \source\ this is perhaps the
best of the Wolf-Rayet candidates, although it is near
the edge of the error circle. 

\medskip

{\it 8. RX J2016.6+3705\/}:  This is a G star of magnitudes $R=11.3$ and
$B=12.4$ that has a typical X-ray flux for its spectral type.

\medskip

{\it 9. RX J2017.6+3637\/}: This is a K star of magnitude $R=11.2$
that has a typical X-ray flux for its spectral type.

\medskip

{\it 15. RX J2016.8+3657\/}: This late K star shows a hint of H$\alpha$
emission in its spectrum, which may be associated with enhanced coronal
X-ray emission.

\medskip

{\it 16. HD 228600\/}:  This is a star of unknown spectral type
with magnitudes $B=10.49$, $V=10.12$. 

\medskip

{\it 17. HD 192641\/}: This is WR 137, a binary of type WC7+OB.  It is
a weak X-ray source, which we list in Table~1 because it was also
suggested as a possible WR counterpart of \source\ by Kaul \& Mitra (1997)
and Romero et al. (1999).

\medskip

{\it 18. HD 228860\/}:  This member of the Cyg OB1 association is of spectral
type B0.5 IV (Humphreys 1978).
It is almost certainly the X-ray source at the eastern edge
of the EGRET error circle, even though its position is poorly determined
this far off axis.  It is variable between two observations separated by
5 months.

\medskip

{\it 19. HD 192639\/}:  The emission lines in this supergiant star,
of spectral type O8 I(f) vary on time scales of several days 
(Rauw \& Vreux 1998).  It is also a variable X-ray source.

\section{Conclusions}

As previously concluded by Mukherjee et al. (2000), we still find that
the most likely identification for \source\ is the blazar \radio.  We
have obtained additional identifications of X-ray sources in the field
to a flux limit of $6.8 \times 10^{-14}$ ergs cm$^{-2}$ s$^{-1}$ for
nearby, relatively unabsorbed sources, or $9.5 \times 10^{-13}$
ergs cm$^{-2}$ s$^{-1}$ for extragalactic or very distant hard X-ray sources.
The SNR CTB 87 is disfavored because of its extreme distance, and 
the fact that energetic pulsars are known to channel only a small fraction,
$< 1\%$ of their power into high-energy $\gamma$-rays.  Cataclysmic
variables, which comprise the majority of the newly identified sources
in this field, are not known $\gamma$-ray emitters.  Similarly, while
there are several Wolf-Rayet and binary O stars in this field,
it remains to be demonstrated that such stars contribute at all
to the EGRET source population.

Our complete set of optical identifications also rules out the alternative
of a Geminga-like pulsar at a distance less than 500~pc, which would
be the maximum distance at which it could be responsible for the
$\gamma$-ray flux of \source.  The \ro\ PSPC flux limit for identified
sources in this field for an assumed blackbody
spectrum of $T = 5 \times 10^5$~K and 
$N_{\rm H} = 3 \times 10^{20}$~cm$^{-2}$
is $2 \times 10^{-13}$ ergs cm$^{-2}$ s$^{-1}$, or 50 times fainter
than Geminga.  Only if $N_{\rm H}$ is as high as
$10^{21}$~cm$^{-2}$ could such a pulsar have escaped detection.
Even then, the $\gamma$-ray properties of \source\ (steep spectrum,
variable) are unlike those of intermediate-age pulsars.

Since there is a bona-fide blazar available for identification with
\source, it remains our candidate of choice.
The probability of finding an EGRET blazar only $1^{\circ}$ from the 
Galactic equator can be estimated from the total number of relatively
well identified blazars, 66, in the Third EGRET Catalog
(Hartman et al. 1999).  This implies an expectation of just one blazar
within the zone $-1^{\circ} < b < +1^{\circ}$.  Thus, we should not
be surprised to have found this one, but we should not expect that blazars
will make a significant contribution to the low-latitude EGRET population.

\acknowledgements
 
We acknowledge the support of NASA grants NAG 5--3696 (RM), 
NAG 5--7935 (EVG), and NAG 5--9095 (JPH). 
This research has made
use of data obtained from HEASARC at Goddard Space Flight Center and 
the SIMBAD astronomical database.

\clearpage

\begin{deluxetable}{lrlcrrlrl}
\tablenum{1}
\tablecolumns{9}
\tablewidth{0pc}
\tablecaption{\ro\ HRI X-ray Sources in the Field of \source\ }
\tablehead
{
    & X-ray & Position &       &        & Optical & Position &     &   \\
No.\tablenotemark{a} & R.A.  & Decl.    & Unc.\tablenotemark{b} & Counts & R.A.    & Decl.    & \omit \hfil $R$ \hfil & \omit \hfil ID  \hfil \\
& (h \hskip 0.5em m \hskip 0.5em s) & ($\circ\ \ \prime\ \ \prime\prime$)  & ($\prime\prime$)
& (ksec$^{-1}$) & (h \hskip 0.5em m \hskip 0.5em s) & ($\circ\ \ \prime\ \ \prime\prime$) & (mag) \\
}
\startdata
 1\tablenotemark{c}
   & 20 16 08.44 & +37 11 28.0 & $\pm 11.$ & $19.2 \pm 3.2$ &              &             &      & CTB 87    \\
 2 & 20 15 36.63 & +37 11 25.1 &  $\pm 0.7$ &  $4.7 \pm 0.4$ &  20 15 36.98 & +37 11 23.2 & 17.5 & CV	      \\
 3 & 20 15 28.35 & +37 11 01.6 &  $\pm 0.7$ &  $4.0 \pm 0.4$ &  20 15 28.76 & +37 10 59.9 & 21.8 & B2013+370 \\
 4 & 20 15 35.71 & +37 04 59.0 &  $\pm 0.7$ &  $7.3 \pm 0.5$ &  20 15 36.10 & +37 04 56.5 & 18.6 & CV        \\
 5 & 20 15 14.87 & +36 59 22.2 &  $\pm 2.2$ &  $2.4 \pm 0.7$ &  20 15 14.73 & +36 59 24.4 & 18.3 & CV        \\
 6\tablenotemark{c}
   & 20 17 28.94 & +37 18 27.0 &  $\pm 10.$ & $31.1 \pm 3.6$ &  20 17 29.70 & +37 18 31.1 &  9.3 & HD 228766 \\
 7 & 20 16 59.77 & +37 25 26.5 &  $\pm 1.3$ & $14.2 \pm 1.1$ &  20 17 00.03 & +37 25 23.8 &  8.1 & HD 193077 \\
 8 & 20 16 37.37 & +37 05 55.4 &  $\pm 1.2$ &  $3.2 \pm 0.4$ &  20 16 37.55 & +37 05 55.0 & 11.3 & G Star    \\
 9\tablenotemark{c}
   & 20 17 36.05 & +36 37 56.1 &  $\pm 10.$ & $34.1 \pm 2.5$ &  20 17 35.86 & +36 38 02.3 & 11.2 & K Star    \\
15\tablenotemark{c}
   & 20 16 48.94 & +36 57 47.1 &  $\pm 6.8$ &  $3.0 \pm 1.2$ &  20 16 49.00 & +36 57 47.9 & 14.2 & K Star    \\
16 & 20 15 30.62 & +37 20 06.6 &  $\pm 2.0$ &  $1.4 \pm 0.4$ &  20 15 30.78 & +37 20 03.1 & 10.8 & HD 228600 \\
17\tablenotemark{c}
   & 20 14 31.94 & +36 39 40.1 &  $\pm 2.0$ &  $6.6 \pm 0.9$ &  20 14 31.77 & +36 39 39.6 & 8.0 & HD 192641 \\
18\tablenotemark{c}
   & 20 18 52.50 & +36 57 43.0 &  $\pm 10.$ &  $21.7 \pm 3.0$ &  20 18 51.55 & +36 57 41.4 & 10.7 & HD 228860 \\
19\tablenotemark{c}
   & 20 14 30.00 & +37 21 15.0 &  $\pm 10.$ &  $15.7 \pm 6.3$ &  20 14 30.43 & +37 21 13.8 & 7.2 & HD 192639 \\
\enddata
\tablenotetext{a}{ The source numbers follow those of Mukherjee et al. (2000).}
\tablenotetext{b}{ 90\% Confidence uncertainty in each coordinate.}
\tablenotetext{c}{ X-ray source data from the \ro\ PSPC.  We assume a positional uncertainty
of $10^{\prime\prime}$ in cases where no value was supplied by the standard analysis.}
\end{deluxetable}
\clearpage

\begin{figure}
\plotone{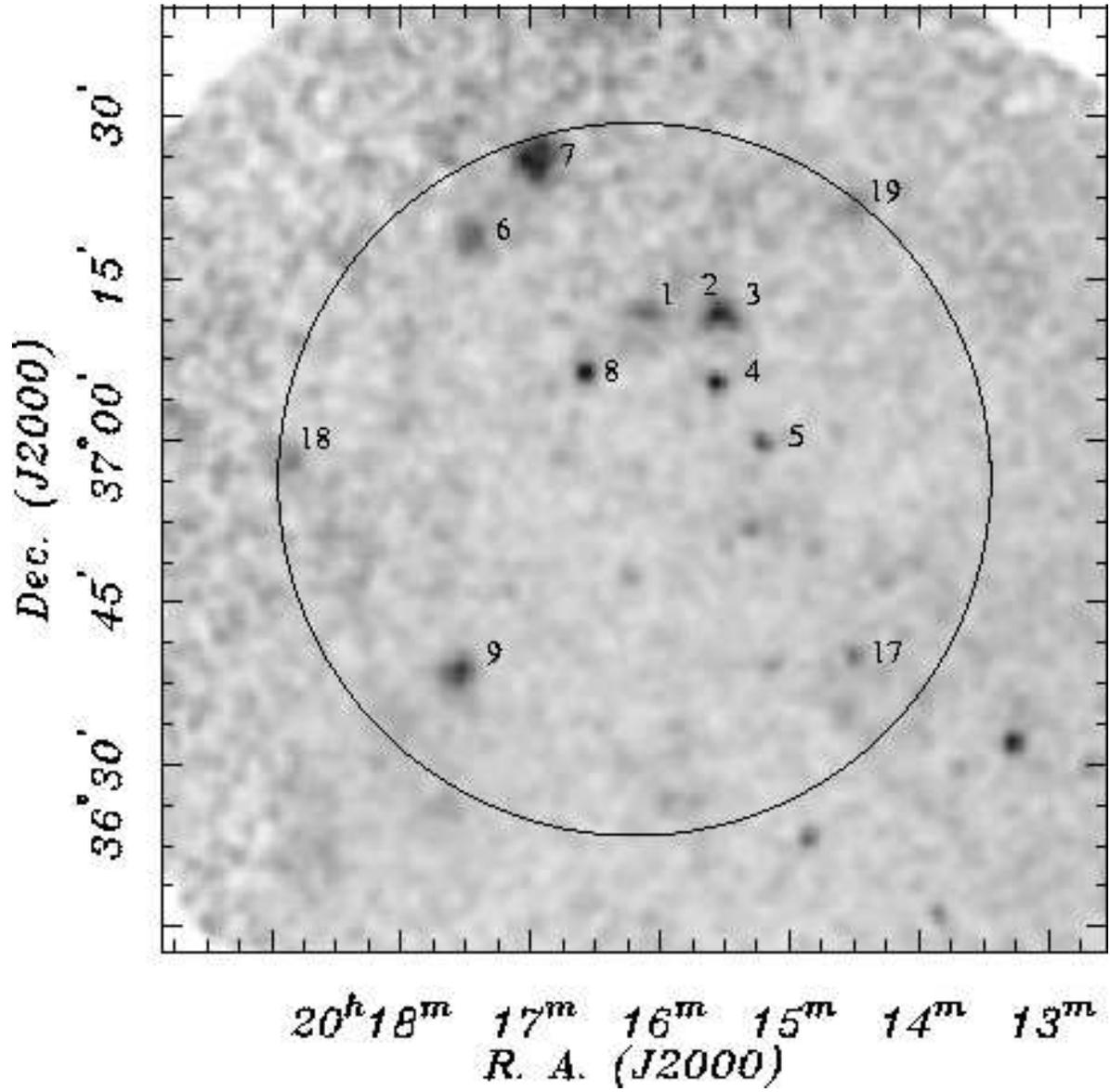}
\figcaption{\ro\ PSPC image and 95\% confidence error circle of
\source, taken from Mukherjee et al. (2000).  The X-ray sources 
are numbered as in that paper and in Table~1.  All sources are point-like
except for \#1, which is the SNR CTB 87.\label{fig1}}
\end{figure}

\begin{figure}
\plotone{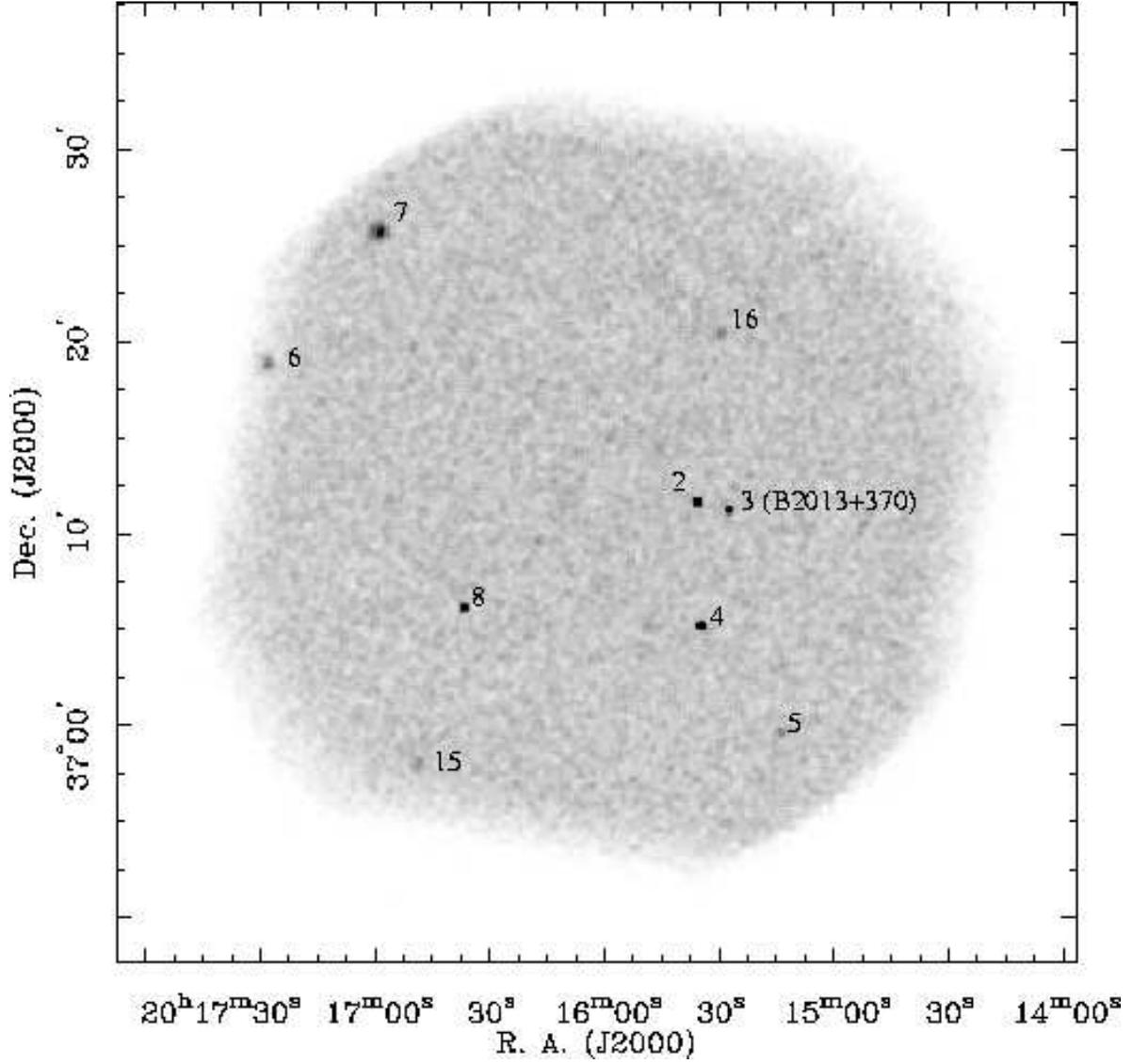}
\figcaption{\ro\ HRI image covering part of the 95\% confidence error 
circle of \source, taken from Mukherjee et al. (2000).  The X-ray sources 
are numbered as in Figure~1.\label{fig2}}
\end{figure}

\begin{figure}
\vskip 2.0 truein
\centerline{THIS FIGURE IS TOO LARGE FOR ASTRO-PH}
\figcaption{Finding charts from the Digitized Palomar Observatory
Sky Survey II red plates for the bright stars which are X-ray sources in
the field of \source.
Each chart is $4^{\prime}\!.3 \times 4^{\prime}\!.3$.
North is up and east is to the left.\label{fig3}}
\end{figure}
\clearpage

\begin{figure}
\vskip -0.5 truein
\plotone{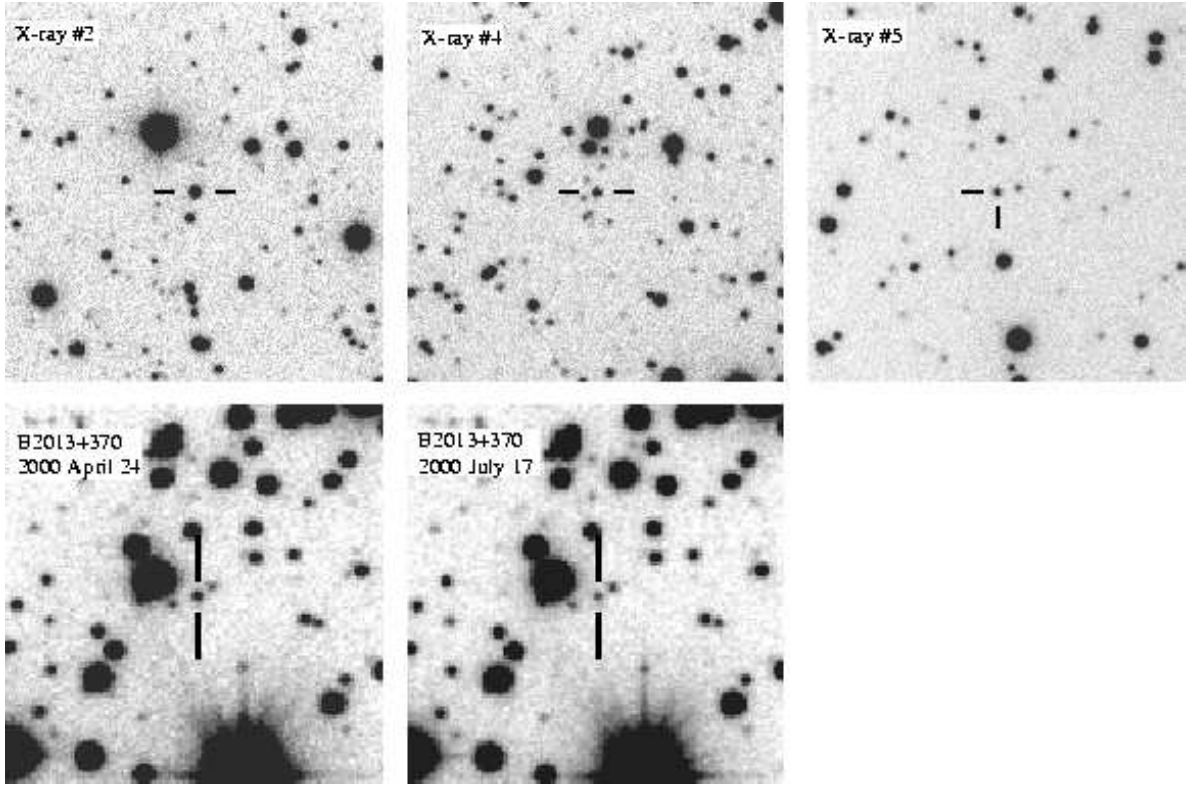}
\figcaption{$R$-band CCD images of faint objects which are
counterparts of X-ray sources in the field of \source.  
The top three images are $70^{\prime\prime} \times 70^{\prime\prime}$,
and the magnitudes of their cataclysmic variable identifications
are given in Table~1.  The two images of the blazar B2013+370 are each
$35^{\prime\prime} \times 35^{\prime\prime}$, and illustrate its change
from $R=21.40 \pm 0.04$ on 2000 April 24 to $R = 21.81 \pm 0.05$ on 2000
July 17.
North is up and east is to the left. 
\label{fig4}}
\end{figure}
\clearpage

\begin{figure}
\plotone{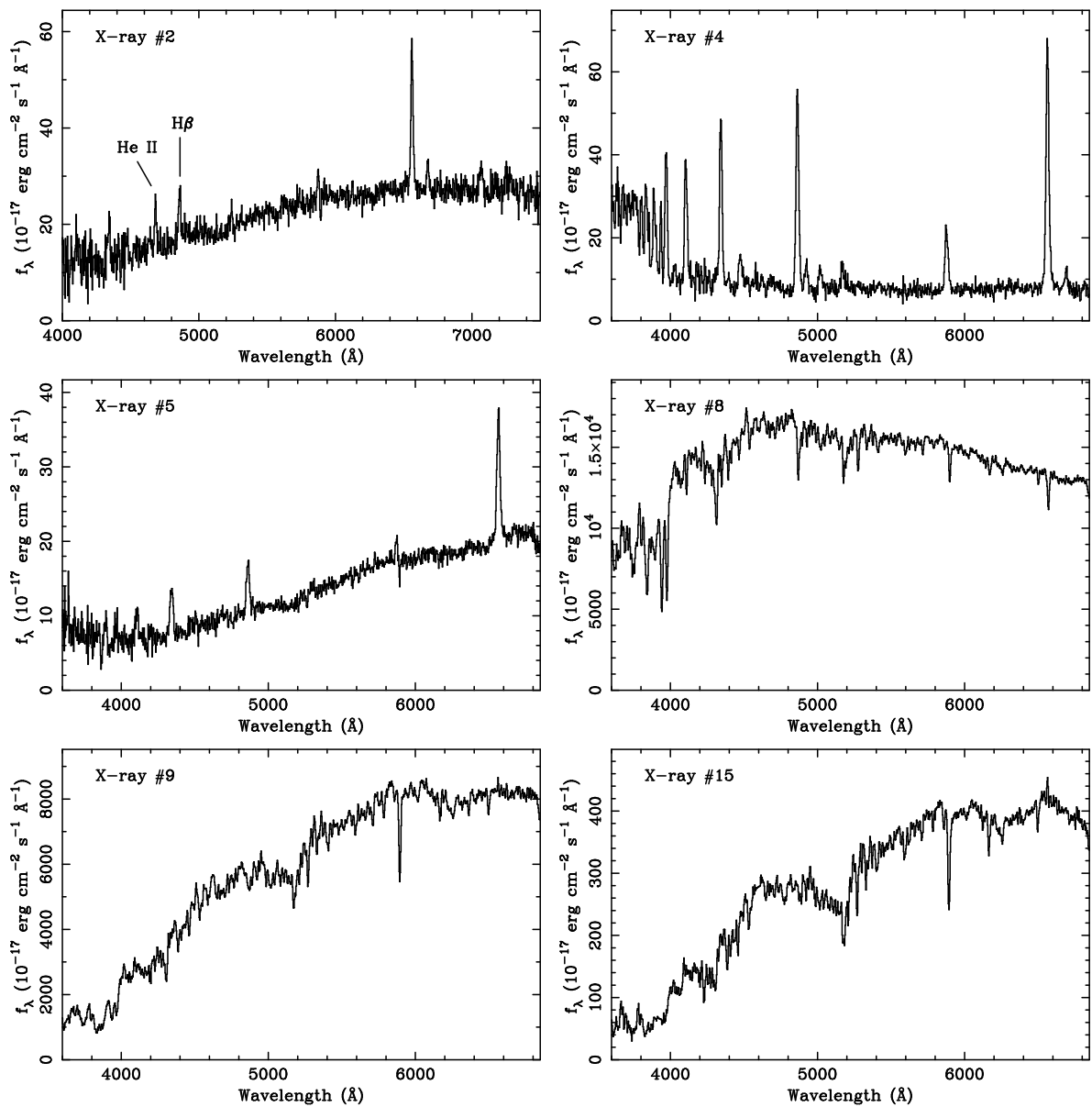}
\figcaption{Spectra of the six fainter optical counterparts
of X-ray sources in the field of \source, obtained on either
the MDM 2.4 telescope or the KPNO 2.1m telescope.\label{fig5}}
\end{figure}
\clearpage

\begin{figure}
\plotone{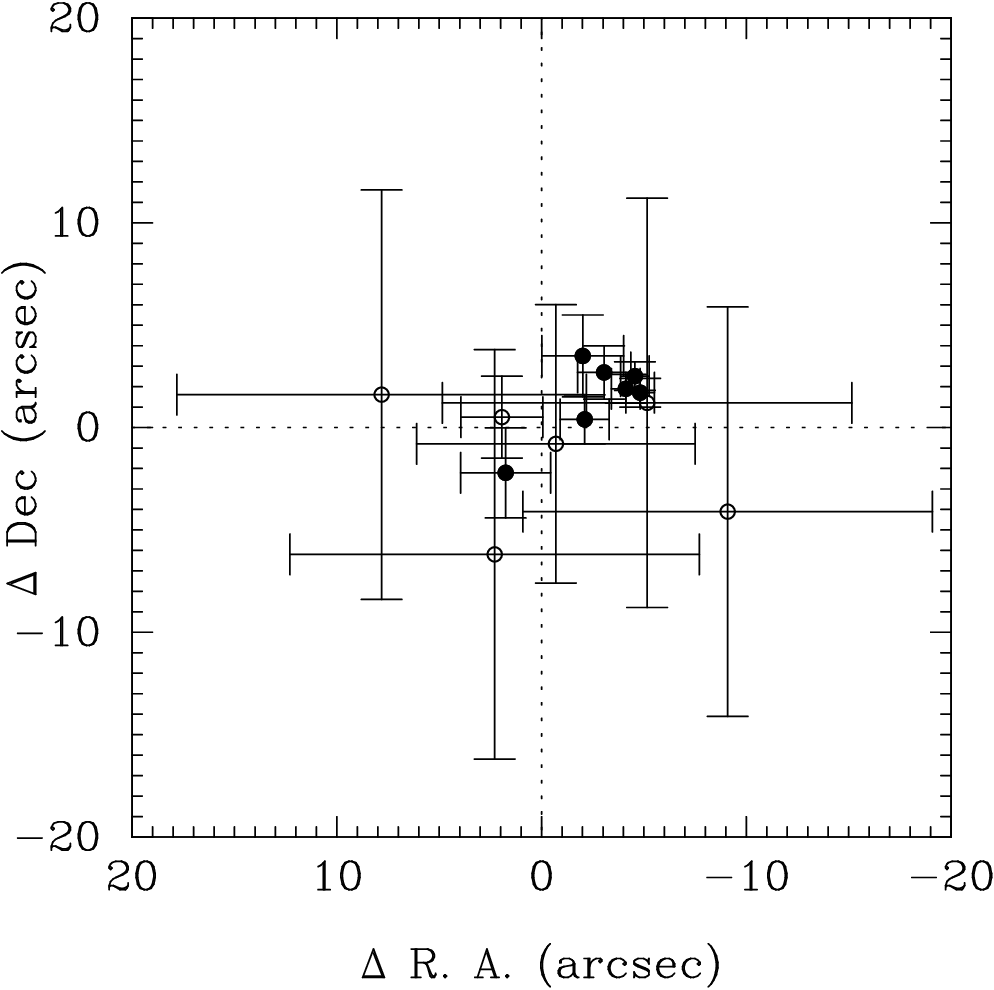}
\figcaption{Offsets between the X-ray and optical positions of
the 13 point-like X-ray sources in Table~1.  Filled circles and
their associated 90\% confidence error bars are from the
\ro\ HRI; open circles and their $1\sigma$ error bars are from the
\ro\ PSPC.\label{fig6}}
\end{figure}
\clearpage

\end{document}